\documentclass[apj]{emulateapj}

\shorttitle{Outflow Mechanisms in Powerful Radio Galaxies}
\shortauthors{Batcheldor et al.}

\begin{document}


\title{Dominant Nuclear Outflow Driving Mechanisms in Powerful Radio Galaxies\altaffilmark{1}}

\author{Dan Batcheldor\altaffilmark{2}, Clive Tadhunter\altaffilmark{3}, Joanna Holt\altaffilmark{3}, Raffaella Morganti\altaffilmark{4}, 
Christopher P. O'Dea\altaffilmark{5},\\ David J. Axon\altaffilmark{6} \& Anton Koekemoer\altaffilmark{7}}


\altaffiltext{1}{Based on observations made with the NASA/ESA {\it Hubble Space Telescope} obtained at the Space Telescope Science Institute, which 
		 is operated by the Association of Universities for Research in Astronomy, Incorporated, under NASA contract NAS 5-26555. These 
		 observations are associated with programme 10206.}
\altaffiltext{2}{Assistant Research Scientist, Center for Imaging Science, Rochester Institute of Technology, 54 Lomb Memorial Drive, 
                 Rochester, NY 14623 \email{dpbpci@astro.rit.edu}}
\altaffiltext{3}{Department of Physics \& Astronomy, University of Sheffield, Sheffield, S3 7RH, UK}
\altaffiltext{4}{ASTRON, Postbus 2, 7900AA Dwingeloo, The Netherlands}
\altaffiltext{5}{Department of Physics, Rochester Institute of Technology, Rochester, NY 14623}
\altaffiltext{6}{Space Telescope Science Institute, 3700 San Martin Drive, Baltimore, MD 21218}


\begin{abstract}
In order to identify the dominant nuclear outflow mechanisms in Active Galactic Nuclei, we have undertaken deep, high resolution 
observations of two compact radio sources (PKS 1549-79 and PKS 1345+12) with the Advanced Camera for Surveys (ACS) aboard the 
{\it Hubble Space Telescope}. Not only are these targets known to have powerful emission line outflows, but they also contain all the 
potential drivers for the outflows: relativistic jets, quasar nuclei and starbursts. ACS allows the compact nature ($<$0\farcs15) of 
these radio sources to be optically resolved for the first time. Through comparison with existing radio maps we have seen consistency in 
the nuclear position angles of both the optical emission line and radio data. There is no evidence for bi-conical emission line features 
on the large-scale and there is a divergance in the relative position angles of the optical and radio structure. This enables us to 
exclude starburst driven outflows. However, we are unable to clearly distinguish between radiative AGN wind driven outflows and outflows
 powered by relativistic radio jets. The small scale bi-conical features, indicative of such mechanisms could be below the resolution 
limit of ACS, especially if aligned close to the line of sight. In addition, there may be offsets between the radio and optical nuclei 
induced by heavy dust obscuration, nebular continuum or scattered light from the AGN. 
\end{abstract}




\section{Introduction}

Gaseous nuclear outflows are known to be present in many Active Galactic Nuclei (AGN). Indeed, these high velocity phenomena have been 
noted in a broad range of objects, including Seyferts, starbursts, quasars and radio galaxies \citep{cren00,kron03,veil04,morg05}. In 
addition, it has been suggested that these outflows can (a) have a significant impact on the surrounding warm inter-stellar 
medium (ISM), which may in turn influence the evolution of the host galaxies, and can (b) limit the growth of the supermassive black 
holes and the rate of star formation \citep{sandr98,wandl03}. As these factors link nuclear activity with the surrounding host galaxy and 
the process of galactic evolution, feedback effects associated with these outflows may be directly responsible for the observed tight 
relations between black hole mass and bulge properties \citep{fandf05}. However, there is a deficiency in our current understanding of 
AGN-induced feedback mechanisms. While the primary source for the large-scale X-ray halos surrounding radio sources, in addition to the 
kilo-parsec-scale radio structures, is seen to be the AGN-induced radio jet in many cases \citep[e.g.,][]{car94,nul02,fab03,gal06}, the 
dominant driving device for these nuclear ($\le 1$ kpc) outflows is still unclear. 

As radio-emitting jets can couple with the ISM more strongly than AGN photons, radio galaxies and radio-loud quasars offer the 
chance to scrutinize the role of nuclear activity on the host galaxies of powerful AGN. In addition, since the light from bright quasar 
nuclei is extinguished along our line of sight by circumnuclear material, we can directly investigate the structure and spatial extent of 
the nuclear outflow regions. Gaseous outflows in powerful radio galaxies are also expected to be important for several reasons. Firstly, 
nuclear activity triggers, i.e., mergers or cooling flows \citep{heck86,bfandc97}, are likely to leave debris which is dissipated as the 
radio source evolves. Such nuclear debris has been implied by the greater number of absorption line systems in young, compact quasars 
than in quasars with extended, evolved radio morphologies \citep{bak02}. Secondly, NICMOS observations of Cygnus A have provided evidence 
of outflow induced despoliation in the form of hollowed out bi-conical structures \citep{tad99}. Finally, it is likely that all the 
suggested drivers of the outflows are present to varying degree in radio galaxies.

\begin{deluxetable}{ccccccccc}
\tabletypesize{\footnotesize}
\tablecaption{Summary of Observations \label{tab:obs}}
\tablewidth{0pt}
\tablehead{
\colhead{Target} &\colhead{Instru.} &\colhead{Filter} & \colhead{Central $\lambda$} & \colhead{Exposures} \\ 
\colhead{}       &\colhead{}        &\colhead{}       & \colhead{${\rm\AA}$}       & \colhead{No. x T(s)} \\
}
\startdata
PKS 1549-79      & WFC1             & FR647M          & 6798                       & 4x250               \\
                 &                  &                 & 7560                       & 4x250               \\
                 & HRC              & FR459M           & 5234                       & 4x730               \\
                 &                  & F550M          & 5580                       & 4x700               \\
                 &                  &                 &                            &                     \\
PKS 1345+12      & WFC1             & FR647M          & 6618                       & 4x200               \\
                 &                  &                 & 7361                       & 4x200               \\
                 & HRC              & FR459M           & 5093                       & 4x640               \\
                 &                  & F550M          & 5580                       & 4x620               \\
\enddata
\tablecomments{Exposures are giving in terms of total number, and length (in seconds), of each drizzle. 
}
\end{deluxetable}

There are several potential sources for the observed nuclear outflows: radiative winds from highly luminous quasar nuclei \citep{bk93}; 
starburst driven winds \citep{ham90}; powerful relativistic jets \citep{bdo97}; cloud interaction with expanding radio lobes 
\citep{ode02}. Each of these mechanisms will leave an imprint on the nuclear morphology of the host. 

The main observational difference between AGN and starburst driven outflows will be in the size of the region from where the process 
originates. Starburst winds will have an origin which is roughly the size of the starburst region, i.e., 100 -- 2000 pc \citep{ham90}, 
whereas the AGN wind driven outflows will have an origin which is inside the torus \citep[e.g.,][]{kandb86}. Therefore, collimated 
outflows on scales of $<100$ pc would indicate that the outflow is AGN driven. A secondary diagnostic is the alignment of the outflow with 
the minor axis of the galaxy. Since starburst driven winds are collimated by the galaxy pressure gradient, they should align with the 
minor axis. However, the torus that defines the AGN driven outflows does not necessarily have to be aligned with the galaxy minor axis as 
Seyfert radio sources show a roughly random distribution of orientation with respect to the minor axis \citep[e.g.,][]{kin00}.

In jet driven outflows the emission line structure will have similar features and scales to the radio structure. However, 
in lobe driven outflows, where the expanding radio lobe bow shock runs over ambient clouds giving them a kick outward, the cloud 
velocities are much smaller than the radio source bow shock expansion velocity. Therefore, the outflow region will be smaller than the
total radio source extent.

The following analysis is designed to determine the dominant mechanism driving nuclear outflows through the study of two compact radio 
sources. For an extended review of such objects see \citet{odea98}. Firstly, the nearby compact flat-spectrum radio source PKS 1549-79 
(z=0.152). Secondly, the nearby Gigahertz Peaked Spectrum (GPS) radio source PKS 1345+12 (z=0.122). Both these sources are hosted by 
ultra-luminous infra-red galaxies (URLIGS), with on-going star formation, and are well known powerful radio galaxies in which outflows 
have been unequivocally detected in the near nuclear narrow-line region, i.e., \citet[][hereafter HTM03]{htandm03} and 
\citet[][hereafter H06]{hol06}. In addition, these previous spectroscopic observations indicate that the [OIII] emission in the nuclear 
regions is dominated by broad, outflowing emission line gas.  Therefore, despite accepting all [OIII] emission from the nuclear regions, 
suitable filters will avoid any ambiguities associated with possible contamination by non-outflowing, narrower emission line components 
in the nuclear regions.

Important information on the outflow driving mechanism could be gained through further consideration of the internal nuclear kinematics 
and also from studying the possible ionization mechanisms. The relative velocities and FWHM of the kinematical components in both 
PKS 1549-79 and PKS 1345+12 are consistent with shocks, however, the same cannot be said about the ionization mechanism. In the case of 
PKS 1549-79, the line ratios are consistent with AGN photoionization; there is no evidence for shock ionization. In PKS 1345+12, the 
nature of the outflow components change between different lines making it impossible to investigate the ionisation mechanisms using the 
standard line ratios and diagnostic diagrams. 

\begin{figure*}[t]
\plotone{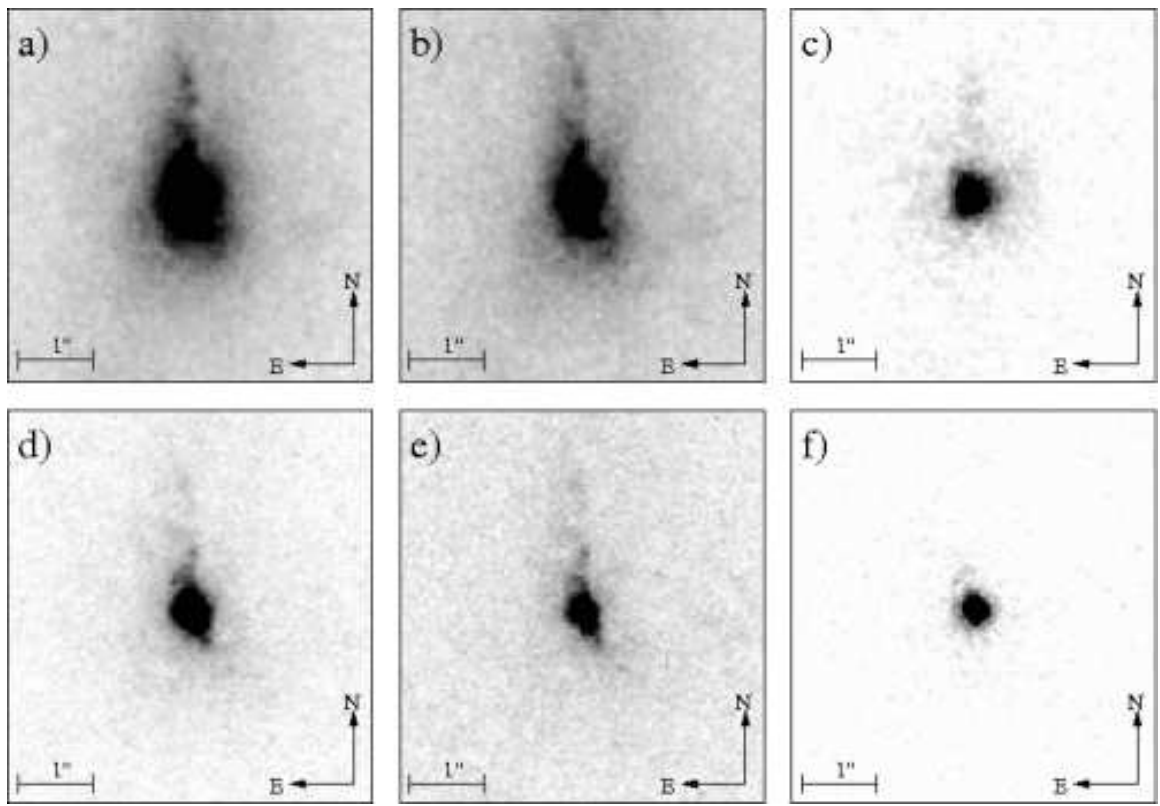}
\caption[PKS 1549-79 ACS Data]{
PKS 1549-79. [{\it Top Row}] WFC1 ACS Observations. (a) FR647M at $7560{\rm\AA}$ (H$\alpha$). (b) FR647M at $6798{\rm\AA}$ (Continuum). 
(c) Continuum subtracted H$\alpha$. [{\it Bottom Row}] HRC ACS Observations. (d) FR550M at $5580{\rm\AA}$ ([OIII]). (e) FR459 at 
$5234{\rm\AA}$ (Continuum). (f) Continuum subtracted [OIII].}\label{fig:1549acs}
\end{figure*}

\begin{figure}[t]
\plotone{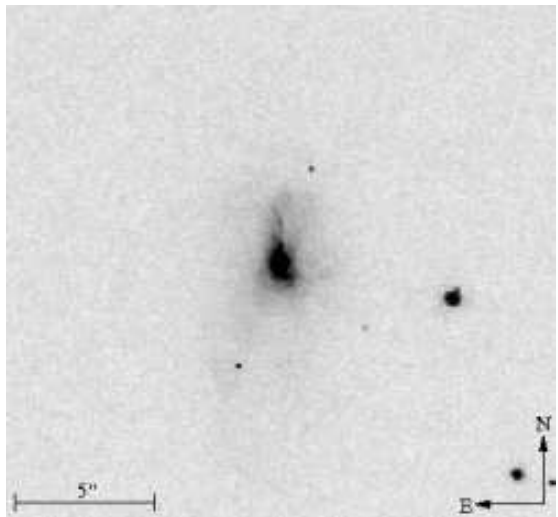}
\caption[PKS 1549-79: Large scale structure]{
PKS 1549-79. The larger structure (FR647M at $6798{\rm\AA}$) showing extended diffuse with potential star clusters and foreground stars.
}\label{fig:1549ls}
\end{figure}
  
We introduce high resolution emission line maps obtained with the Advanced Camera for Surveys (ACS) aboard the {\it Hubble Space 
Telescope} ({\it HST}). The superior spatial resolution of ACS allows the compact nature of these sources ($\sim$0\farcs15) to be resolved 
for the first time. These new data make it possible to distinguish the main outflow driving mechanism through morphological comparison 
with existing radio data. 
  
In \S~\ref{obs+dr} we detail the {\it HST} observations and the data reduction steps. The new ACS data are presented in \S~\ref{1549} and 
\S~\ref{1345}. These observations, in conjunction with VLBI data, allow us to attempt to identify the mechanisms dominating the nuclear 
outflows in \S~\ref{dommech}. Our findings are discussed in \S~\ref{discussion} and concluded in \S~\ref{cons}.

\section{Observations and Data Reduction}\label{obs+dr}

PKS 1549-79 and PKS 1345+12 were observed by ACS on the $4^{\rm th}$ and $5^{\rm th}$ August 2004 as part of Cycle 13 \#10206 
PI: Tadhunter. Two exposures of PKS 1549-79, each of 4x250s, were made using chip 1 of the Wide Field Channel (WFC1, pixel scale 
$\approx0\farcs05$ -- the spatial scale at the distance of these targets is $\sim~3{\rm~kpc/arcsec}$)\footnote{We adopt a value of 
$\rm H_0~=~70~km~s^{-1} Mpc^{-1}$ throughout}. The IRAMP (FR647M, $\delta\lambda=9\%$) ramp filter was adjusted to cover two central 
wavelengths, $7560{\rm\AA}$ for the H$\alpha$ emission line and $6798{\rm\AA}$ for the continuum map. To avoid overfilling the {\it HST} 
storage buffer only a 1024x1024 sub-array of the WFC1 was read out. Two more exposures of PKS 1549-79 were also made using the High 
Resolution Channel (HRC, pixel scale $\approx0\farcs027$). The first focused on the [OIII] emission line (F550M, 
$\delta\lambda=547{\rm\AA}$, 4x700s) and the second focused on the continuum (FR459M - $5234{\rm\AA}, \delta\lambda=9\%$, 4x730s). This 
strategy was repeated for PKS 1345+12 where the central wavelength of the FR647M ramp filter was adjusted to $7361{\rm\AA}$ and 
$6618{\rm\AA}$ (4x200s each), and $5093{\rm\AA}$ for the FR459M ramp filter (4x640s). A summary of these observations is presented in 
Table~\ref{tab:obs}.

\begin{figure*}[t]
\plotone{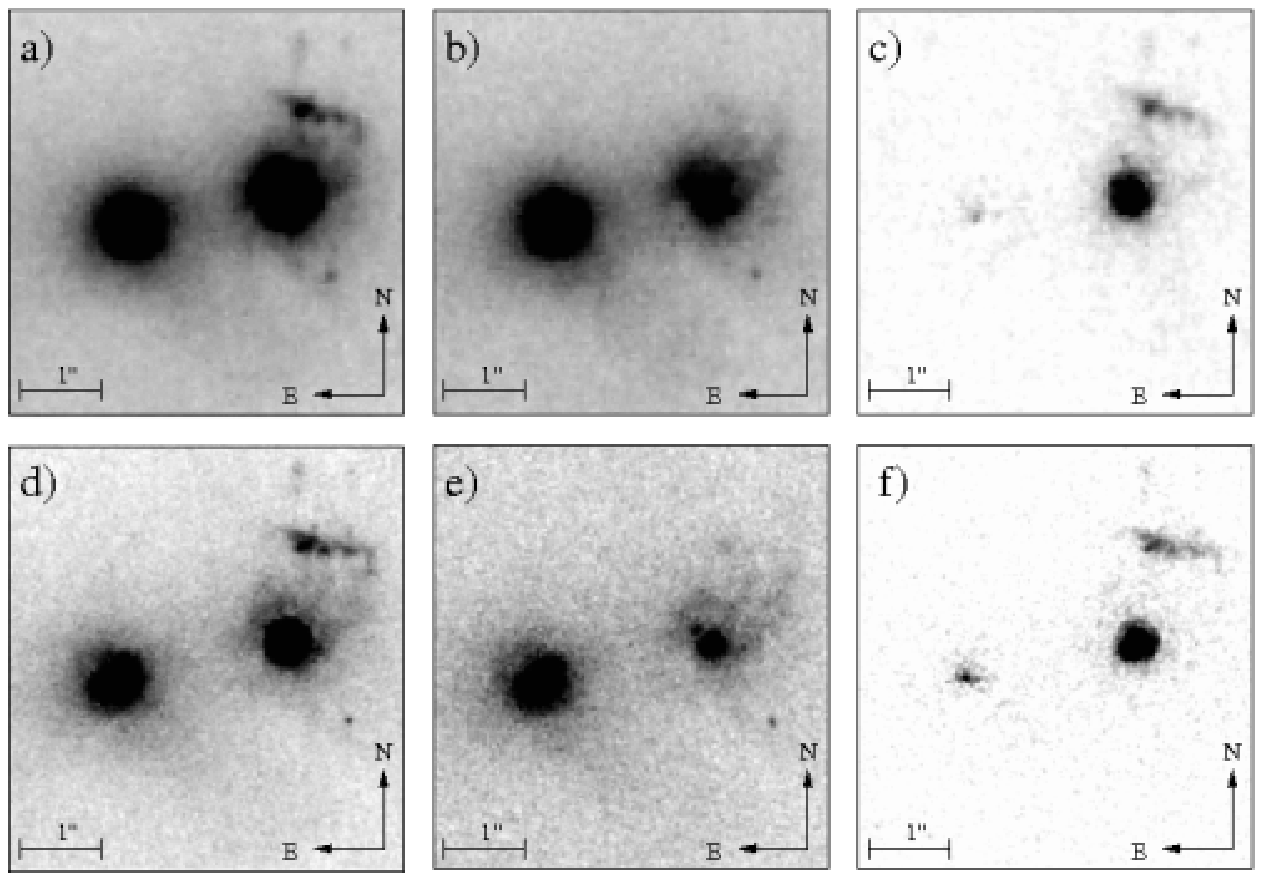}
\caption[PKS 1345+12 ACS Data]{
PKS 1345+12. [{\it Top Row}] WFC1 ACS Observations. (a) FR647M at $7361{\rm\AA}$ (H$\alpha$). (b) FR647M at $6618{\rm\AA}$ (Continuum). 
(c) Continuum subtracted H$\alpha$. [{\it Bottom Row}] HRC ACS Observations. (d) FR550M at $5580{\rm\AA}$ ([OIII]). (e) FR459 at 
$5093{\rm\AA}$ (Continuum). (f) Continuum subtracted [OIII].}\label{fig:1345acs}
\end{figure*}

\begin{figure}[t]
\plotone{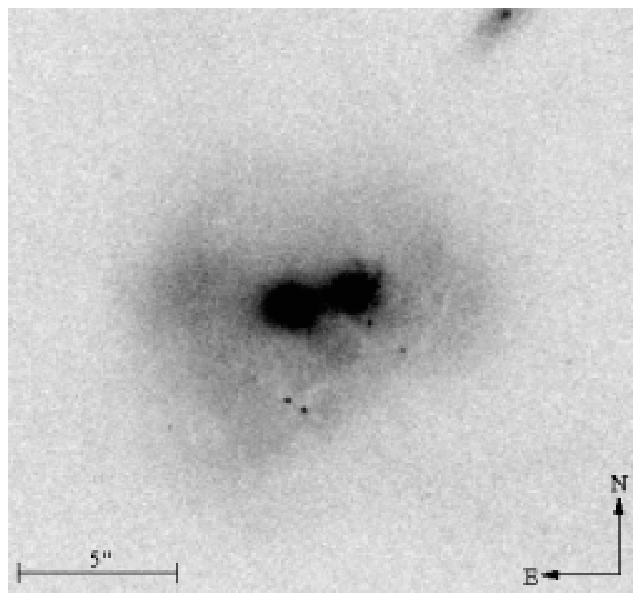}
\caption[PKS 1345+12: Large scale structure]{
PKS 1345+12: The larger scale structure (FR647M at $6618{\rm\AA}$) showing the distorted morphology.
}\label{fig:1345ls}
\end{figure}

The data were passed through the ACS reduction pipeline, using the latest reference files, in order to remove instrumental residuals. 
Cosmic-ray removal was facilitated through the use of the Multi-drizzle script \citep{koek02}. Detailed examinations of the distortion 
corrected data were performed, given that only subsets of the data were read out. The ellipticity of isophotes fitted to 10 stellar 
objects across each field never exceeded 0.08 and averaged 0.03. As a final step, the on and off line images were scaled and subtracted 
from each other in order to produce emission line maps.

\section{PKS 1549-79}\label{1549}

The radio source PKS 1549-79 has had much previous work carried out on it. A brief outline of its main properties will be given before 
the new data are presented. Readers interested in more thorough reviews are directed toward \citet{morg01}, \citet{tad01} and H06.

VLBI observations reveal PK1549-79 to have a relatively small ($\sim$150 mas) one-sided distorted jet structure bending through 60\degr 
\citep{king94}. Lower resolution maps show no evidence of large scale structures. The radio jet has a steep radio spectrum and originates 
from an unresolved flat spectrum radio core. Although the jet is likely close to our line of sight this source shows no evidence for broad 
emission-line features, contrary to the standard orientation based unification schemes. A model describing a cocooned (obscured) quasar, 
where a young radio jet is carving a path through a dense nuclear region, has been proposed by \citet{tad01}. A narrow HI absorption line 
gives a redshift of 0.152 \citep[][H06]{morg01}. \citet{pandp83} were first to associate the radio source with what has been classified 
as an 18.5(V) magnitude Seyfert 2. 

Signs of star formation are provided by abnormally strong far-infrared (FIR) emissions, i.e., ULIRG, and an optical continuum that is 
dominated by a population of early type stars \citep{randn97,dick97}. Based on spectral synthesis modeling, H06 have shown this young 
stellar population to account for 1-30\% of the total stellar mass, and be between 50 and 250 Myrs old. In addition, the modeling of H06 
has shown that the reddened quasar, which has been directly detected in the K-band \citep{bell03}, makes a small but significant 
contribution to the continuum around the wavelength of H$\alpha$. This has also been suggested by the detection of a very broad component 
to the H$\alpha$. Further spectral analyses \citep[][H06]{tad01} have shown the outflow to have an extremely broad high ionization [OIII] 
line (FWHM$\sim$1280${\rm~km~s^{-1}}$), and a lower ionization [OII] line (FWHM$\approx$380${\rm~km~s^{-1}}$). The outflow, which is 
blue-shifted by $680{\rm~km~s^{-1}}$ with respect to the rest frame, is detected in all of the ionization lines. However, the outflow 
is strongest (and dominated by) the high ionization lines. In comparison, the low ionisation lines are dominated by quiescent gas.
                                      
In summary, H06, using ground-based optical, infrared (IR) and radio observations, have confirmed that PK1549-79 constitutes a young 
luminous narrow-line (obscured) quasar, the activity of which has been triggered by a gas rich major merger. 

\begin{deluxetable}{lccc}
\tabletypesize{\footnotesize}
\tablecaption{Astrometric Details\label{tab:astro}}
\tablewidth{0pt}
\tablehead{
\colhead{} &\colhead{PKS 1549-79} &\multicolumn{2}{c}{PKS 1345+12} \\ 
\colhead{} &\colhead{}            &\colhead{(East)}      &\colhead{(West)}
}
\startdata
{\it HST} RA:  & $15^{\rm h}56^{\rm m}58\fs96$     & $13^{\rm h}47^{\rm m}33\fs51$ & $13^{\rm h}47^{\rm m}33\fs37$ \\
{\it HST} DEC: & $-79\degr14\arcmin04\farcs16$     & $+12\degr17\arcmin22\farcs99$ & $+12\degr17\arcmin23\farcs44$ \\
Radio RA:      & $15^{\rm h}56^{\rm m}58\fs87^{1}$ & \multicolumn{2}{c}{$15^{\rm h}56^{\rm m}58\fs87^{2}$}         \\
Radio DEC:     & $-79\degr14\arcmin04\farcs28^{1}$ & \multicolumn{2}{c}{$-79\degr14\arcmin04\farcs28^{2}$}         \\ 
$\Delta$RA:    & $1\farcs35$                       & $2\farcs25$                   & $0\farcs15$                   \\
$\Delta$DEC:   & $0\farcs12$                       & $1\farcs27$                   & $0\farcs82$                   \\
\enddata
\tablecomments{All co-ordinates are in J2000. (1) Radio data from \citet{john95}. (2) Radio data from \citet{stang01}.
}
\end{deluxetable}

In Figure~\ref{fig:1549acs} we present the new ACS data for PKS 1549-79. In both the H$\alpha$ and [OIII] line and continuum images 
we find a prominent nucleus with an offshoot extending to the north. The radio and {\it HST} astrometry is listed in 
Table~\ref{tab:astro}. The offshoot, which is significantly less prominent in the [OIII] emission, appears to have several distinct 
components, the furthest of which lies along a position angle (PA) of 0 and at a distance of $1\farcs6$ ($\sim5$ kpc) from the 
nucleus. The sharpness of the features in this northern offshoot, and the block-like appearance of the isophotes to the south-east 
(PA$\approx15\degr$) of the nucleus provide evidence for a dust lane with a PA similar to the PA of the nuclear emission line and 
continuum structures (e.g. Figure~\ref{fig:1549acs}d). 

On the large scale (Figure~\ref{fig:1549ls}) we can see extremely faint, diffuse emission that extends for $\sim6\farcs5$ (20 kpc) in 
the east-west direction. The north-south extension, as noted by \citet{pandp83}, can be seen. This Figure shares similar features to the 
deep Very Large Telescope Gunn $r$ image of H06 (their Figure~1). The authors note that the north-south extended features appear to 
contain knots of emission. Here we can see point-like sources in similar positions to these knots, which could be foreground stars but may 
also be super star clusters associated with the tidal tails of PKS1549-79. 
                                                                    
\section{PKS 1345+12}\label{1345}

The radio source PKS 1345+12 has also had much previous work carried out on it. Again, a brief outline of its main properties will be 
given before the new data are presented. Readers interested in a more thorough review are directed toward HTM03. 

PKS 1345+12 (4C 12.50) is one of the closest GPS sources. This allows the compact double jet and core structure, noted from VLBI (2cm) 
and VLBA (6cm) radio observations, to present proportions of $\sim0\farcs15$. The more prominent S-shaped jet component extends to the 
south-east before bending and expanding into a diffuse lobe. Protruding to the north-west of the core (the core being identified from the 
relative flatness of the radio spectrum) weak radio emission is detected. On larger scales, diffuse radio emission is detected extending 
35\farcs0 ($\sim$80 kpc) to the north, and 25\farcs0 ($\sim$55 kpc) to the south \citep{stang05}.

Optically, PKS 1345+12 shows a complex morphology characterized by two nuclei, the western most of which, identified as a 17.5(V) 
magnitude elliptical, possessing an extended curved tail. Previous {\it HST} studies, which are unable to resolve structure down to the 
scale of the radio morphology, have shown the western nucleus to be associated with the radio source \citep{axon00}. From the ACS data 
presented here we find optical -- radio offsets (Table~\ref{tab:astro}) similar to those found by \citet{axon00}, which confirms the 
western nucleus as the host of the radio source. The double nucleus and distorted appearance of PKS 1345+12 shows that a merger event is 
taking place. As with the other source in this study, PKS 1345+12 shows a young stellar population \citep{tad05}, a FIR excess (ULIRG) and 
near-UV emission \citep[][Labiano et al., in prep]{evans99}, indicating the presence of considerable star formation. Using optical 
spectroscopy, HTM03 have observed 3 distinct nuclear kinematical components in PKS 1345+12, the narrowest of which 
(FWHM$\approx340{\rm~km~s^{-1}}$) is interpreted as the systemic velocity. The two other components, designated ``intermediate'' and 
``broad'', show FWHM of $\sim1250{\rm~km~s^{-1}}$ and $\sim1950{\rm~km~s^{-1}}$ respectively, with corresponding blue-shifts of 
$\sim400{\rm~km~s^{-1}}$ and $\sim1980{\rm~km~s^{-1}}$ with respect to the rest frame. Due to the reddening observed in each component, 
it is proposed that the broad component originates from the inner most regions closest to the obscured quasar, whilst the narrow component 
represents a quiescent halo.
 
In Figure~\ref{fig:1345acs} we present the new ACS data for PKS 1345+12. In both the H$\alpha$ and [OIII] line and continuum images we 
find the two separate nuclei clearly visible within an extended diffuse envelope. The $2\farcs0$ ($\sim5$ kpc) separation of the two 
nuclei measured here is $\sim0\farcs1$ greater than the separation reported by \citet{heck86} and \citet{gands86}. The continuum 
subtracted emission line images (Figures~\ref{fig:1345acs}c and f) show that an insignificant amount of line emission emanates from the 
eastern component when compared to the western nucleus. To the north-west of this western nucleus there is an extended emission line 
filament (e.g., Figures~\ref{fig:1345acs}a, c, d and f). Immediately north-east of the main nuclear structure there is a separate island 
of [OIII] emission (see Figure~\ref{fig:1345acs}e) and a ``mushroom'' like morphology to the H$\alpha$ continuum (see 
Figure~\ref{fig:1345acs}b); this is clear evidence of a dust lane running approximately south-east to north-west. On the larger scale we 
can again compare this data to the findings of \citet{heck86}. Figure~\ref{fig:1345ls} highlights the distorted morphology and hints at 
some of the south-west curved tail extensions mentioned by \citet{heck86}. We also see that the halo of PKS 1345+12 contains similar point 
like sources (to the south of the eastern nucleus) to the foreground stars or possible star cluster seen around PKS 1549-79. Further study 
has shown these sources to be super star cluster with young stellar populations \citep{rod06}.

\section{Investigating the Dominant Mechanism}\label{dommech}

In this section we closely examine the inner structures of the observed nuclei and determine the most likely position of the AGN in 
the ACS emission line images (and therefore the likely position of the radio cores). Bi-conical structures will suggest that the outflows 
are wind driven, whereas structures similar to the radio morphologies will suggest that the outflows are driven by the radio jets. 
However, first we examine the relative astrometry in the ACS data in order to estimate the uncertainties in derived positions.

\subsection{Relative Astrometry and Alignments}

There are two elements that could contribute to differences between the spatial offsets of objects measured on the WFC1 and HRC 
frames: the errors in measuring the positions themselves, i.e., true positions as related to measured positions, and the residual 
errors in the correction of the spatial distortion by the pipeline reduction. By quantifying these errors we can estimate the astrometric 
uncertainties present in the ACS images. An estimate of the residual distortion error can be gained from the relative positions of stellar 
like objects present in both the WFC1 and HRC data. The errors in measuring the positions themselves can be minimized by averaging the 
differences across as many objects as possible. We identified 7 stellar like objects common to the PKS 1549-79 data (the PKS 1345+12 field 
is relatively empty) and centroided their positions. Each stellar object was then paired with the remaining six objects, and the offset 
between the pairs were recorded (21 total offsets). The dispersion in the relative offsets between these objects in the WFC1 and HRC 
images, which we adopt as our astrometric uncertainty, was found to be $\sim0\farcs03$.

\begin{figure*}[t]
\plotone{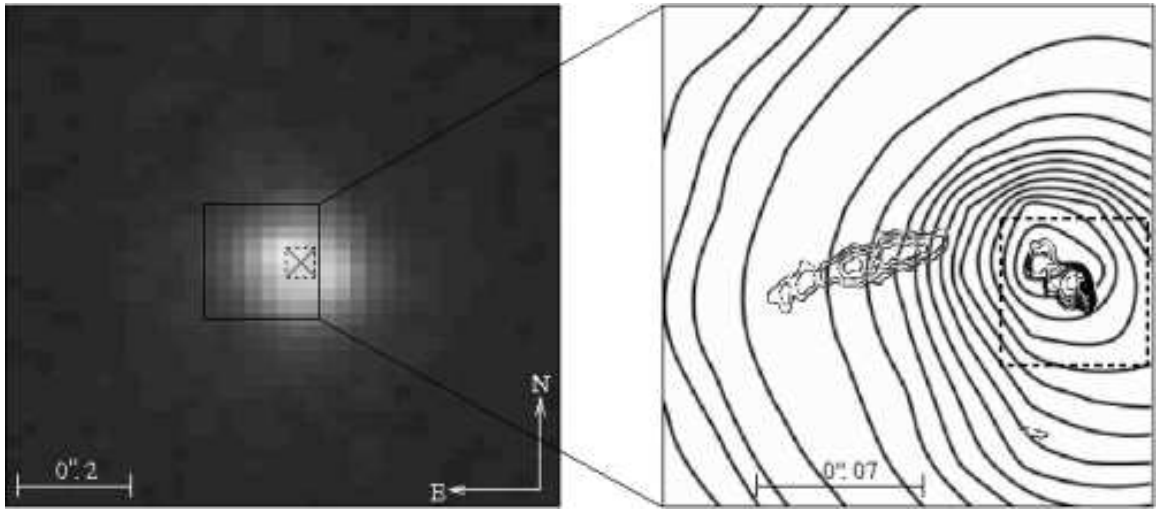}
\caption[PKS 1549-79: Comparing with the radio morphology]{
PKS 1549-79: Comparing the ACS data with the radio morphology. On the left is the HRC continuum subtracted [OIII] emission line map with 
a logarithmic stretch showing the very high surface brightness features. An ``X'' marks the position of the H$\alpha$ core and dotted lines 
show the estimated astrometric error box. On the right we see an expanded view of the [OIII] emission (smoothed thick contours) overlaid 
with the 2.3 GHz radio map (thin contours at 10, 20, 40, 80, 160, 320, 640 and 1280 mJy Beam$^{-1}$).
}\label{fig:1549rad}
\end{figure*}

\begin{figure}[t]
\plotone{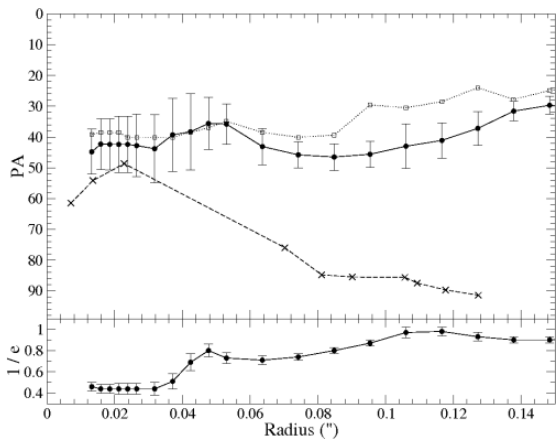}
\caption[PKS 1549-79: Comparing optical and radio position angles]{
PKS 1549-79: Comparing continuum subtracted [OIII] and radio position angles. In the top panel, positions of knots in the radio emission, 
with respect to the radio core, are plotted as crosses and joined with a dashed line, while the PAs of ellipses fitted to the 
[OIII] continuum subtracted emission are joined with a solid line. The PAs of the [OIII] continuum are plotted with open squares and 
joined with a dotted line, the errors are omitted for clarity but are at the same level as for the [OIII] emission. The lower panel 
shows the ellipticities of the [OIII] continuum subtracted emission line data.
}\label{fig:1549pas}
\end{figure}

In addition to the astrometric uncertainties, the obscured nature of the central engines can be an added source of error for the relative 
alignments of the AGN and radio cores; the measured peak in emission line flux may not correspond to the position of the AGN. In the 
following two sections we will attempt to minimize this effect, however, for this process to be a significant source of error it must act 
at a distance of $>0\farcs03$ (100 pc) from the estimated position of the central AGN. 

\subsection{PKS 1549-79}\label{sub1549}

The very central region (1\farcs0) of PKS 1549-79 ([OIII] emission line) is presented in Figure~\ref{fig:1549rad}(a). We see a relatively 
high surface brightness concentration to the north-east, with a lower surface brightness fan of emission extending toward the south-west. 
Based on existing optical and near-IR spectra we do not expect to be able to detect the reddened AGN in the HRC [OIII] images. However, 
we know from the modeling presented by H06 that the quasar nucleus should be detected in the H$\alpha$ continuum image. Therefore, we 
have assumed that the centroid of the concentration of pixels at the center of the WFC1 H$\alpha$ continuum image gives the true quasar 
position relative to the reference stars common to both the WFC1 and HRC fields. When translated to the HRC images the WFC1 position is 
found to lie at the point marked with an ``X'' in Figure~\ref{fig:1549rad}(a). It is with this point that we align the radio core. The 
estimated astrometric uncertainty is outlined with dotted lines ($\pm0\farcs03$).

The radio map in panel (b) of Figure~\ref{fig:1549rad} (thin contours) is that presented by H06.  The data were collected in 1988 at 
$\sim3$ GHz using a the Southern Hemisphere VLBI Experiment, SHEVE \citep{pre84}. The beam size is $7.3\times2.7$ mas at a PA of 4.7\degr. 
The western-most knot shows a flat spectrum across 2.3-8.4 GHz and is considered the core. The H$\alpha$ AGN position has subsequently 
been aligned with this. A steep spectrum jet is seen to extend to the east at a PA of $\sim90\degr$. The jet bends through approximately 
60\degr, indicative of orientation change, and exhibits a gap which may be attributed to cyclic emission processes. H06 estimate an upper 
limit of $i<55\degr$ for the inclination, with respect to the line of sight, for this jet. 

\begin{figure*}[t]
\plotone{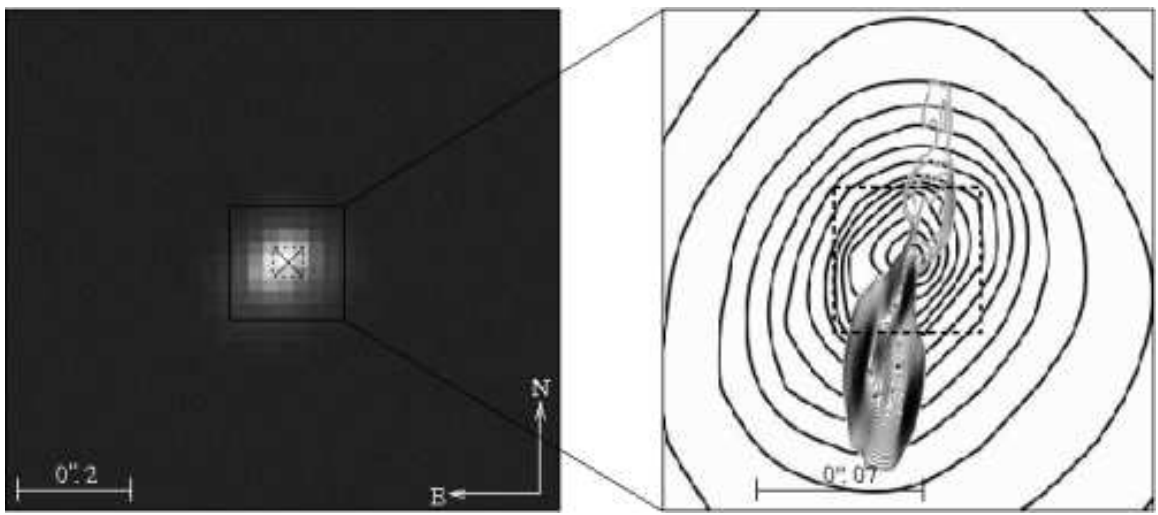}
\caption[PKS 1345+12: Comparing with the radio morphology]{
PKS 1345+12: Comparing the ACS data with the radio morphology. On the left is the HRC continuum subtracted [OIII] emission line map with 
a logarithmic stretch showing the very high surface brightness features. An ``X'' marks the position of the NIR core and dotted lines show 
the estimated astrometric error box. On the right we see an expanded view of the [OIII] emission (smoothed thick contours) overlaid with the 
1.3 GHz radio map (thin contours at 7, 10, 16, 25, 37, 42 ... 766 mJy Beam$^{-1}$).
}\label{fig:1345rad}
\end{figure*}

\begin{figure}[t]
\plotone{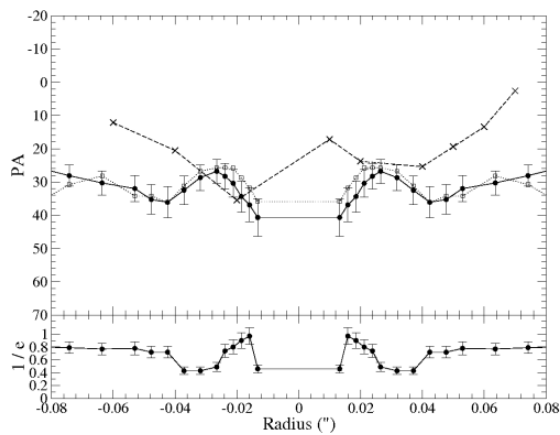}
\caption[PKS 1345+12: Comparing optical and radio position angles]{
PKS 1345+12: Comparing continuum subtracted [OIII], [OIII] continuum and radio position angles. As Figure~\ref{fig:1549pas} but for 
PKS 1345+12.
}\label{fig:1345pas}
\end{figure}

Figure~\ref{fig:1549rad}(b) also demonstrates the comparative morphologies between the radio structure (thin contours) and the [OIII] 
emission (thick lines). The emission line map shows no evidence for bi-conical features. Ellipse fitting to the [OIII] data is presented 
in Figure~\ref{fig:1549pas}, where the PAs of knots along the jet, with respect to the radio core, are also plotted. We see excellent 
agreement between the [OIII] continuum and [OIII] emission line PAs. Similarity in the emission line and continuum morphology could be 
attributed to dust obscuration, scattered AGN light \citep{tad92}, nebular continuum \citep{dic95}, or a mixture of all three. 
In addition to the optical emission-continuum alignment, there is an overlap in the very inner regions between the radio structure and 
the [OIII] emission. However, this radio-optical agreement deteriorates as the jet curves around. The jet also extends past the highest 
surface brightness region of the [OIII] emission. 

\subsection{PKS 1345+12}\label{sub1345}

The very central region of the western nucleus in PKS 1345+12 ([OIII] emission line) is presented in Figure~\ref{fig:1345rad}(a). A 
north-west to south-east elongation is clearly apparent. As in the case of PKS 1549-79, we do not expect to be able to directly detect the 
AGN in the [OIII] images, nor at optical wavelengths. However, we can assume that the quasar nucleus will be detected in the K-band. 
Therefore, we have retrieved a NICMOS F222M image for this source from the {\it HST} 
archive\footnote{\tt http://archive.stsci.edu/hst/search.php}. It was taken as part of programme \#7219 PI: Scoville, the finer details of 
which are presented in \citet{scov00}. By determining the relative offset of the eastern nucleus with that of the western nucleus in both 
the emission line images and the NIR, we can place a better constraint on the position of the radio core. Following this we have indicated 
in Figure~\ref{fig:1345rad}(a) the position of the NIR peak with respect to the eastern nucleus on the HRC emission line map. It is here 
we assume the radio core to fall.

The radio map in panel (b) of Figure~\ref{fig:1345rad} (thin contours) is a 1.3 GHz map obtained from a global VLBI (HI) experiment 
\citep{mor04} overlaid onto the smoothed contours of the HRC emission line map (thick lines). The radio data were collected in 2001 where 
the beam size was $6.6\times1.2$ mas at a PA of -11\degr. The radio core has been identified at 5 GHz by \citet{stang97} from the 
relatively flat spectrum. However, similar to the findings of \citet{xia02}, this core is not clearly defined in the 1.3 GHz map, and is 
likely self absorbed. Therefore, in order to register the 1.3 GHz map (which shows more extended, lower surface brightness features than 
the 5 GHz map) against the optical emission line images, we have aligned the 5 GHz map with the 1.3 GHz map based on the positions of the 
southern components of the jet -- where optical depth effects are likely to be less of an issue. Then, knowing the true position of the 
radio core in the 1.3 GHz map, we can place it onto the emission line image. 

Ellipse fitting to the [OIII] data is shown in Figure~\ref{fig:1345pas}. As radio emission is detected either side of the nucleus, we have 
presented the PA of both the northern and southern jets with respect to the radio core, using knots in both the 1.3 and 5 GHz maps. We can 
see that there is excellent agreement with the radio and [OIII] across a range of radii, but there is significant divergence as the jet 
extends past the high surface brightness regions and begins to bend. As with PKS 1549-79, we can also see the excellent agreement between 
the morphologies of the [OIII] continuum and emission line structures. 
                                                
\section{Discussions}\label{discussion}
                    
There are several features of these high resolution data that are common to both objects. In the first instance, there are strikingly 
detailed similarities in the morphologies of the continuum and emission lines detected in the nuclear regions. For example, in both 
objects the [OIII] continuum images are very similar to the detailed morphologies of the the [OIII] emission in terms of both alignment 
and scale. This is particularly evident from Figures~\ref{fig:1549pas} and \ref{fig:1345pas}. Secondly, in both objects the relative 
alignments in the PAs of the optical morphologies and radio knots follow the same pattern; close alignments in the inner regions and 
increasing disparities at larger radii. In PKS 1549-79 the observed radio core is not exactly aligned with the [OIII] emission, it is 
offset to the south-east. However, considering the estimated positional accuracy, this offset is not significant. We can see from 
Figure~\ref{fig:1549pas} that the nuclear PAs of both the radio and optical data are consistent. In addition, the PAs fitted to the 
nuclear isophotes of PKS 1345+12 compare remarkably well with PAs of the radio jet knots for a radius $<0\farcs04$ 
(Figure~\ref{fig:1345pas}). The isophotes also tentatively follow some of the radio structure, i.e., PAs change in tandem with the jet. In 
both cases, nevertheless, the optical morphologies do not resemble the nature of the jet on larger scales ($>$0\farcs15). This is 
consistent with the idea that the jets are being deflected at the location of the high surface brightness [OIII] emission.

From inspection of Figures~\ref{fig:1549rad} and \ref{fig:1345rad} we find no evidence of bi-conical structures extending beyond the 
radio jets. The ellipticities of isophotes fitted to the extended emission line data are low ($\sim$0.05). On the smallest scales 
($\sim0\farcs07$) there is also no evidence for bi-cones. Adding this observation to the de-coupling of the radio and optical 
morphologies at radii greater than $\sim130$ pc means that we can rule out starburst wind driven outflows; the expected extent of 
starburst winds \citep{ham90} are greater than the scales seen here.
    
The close alignment of PAs between the radio and optical, in the inner regions, provides clear evidence for the relativistic radio jet 
driven outflow mechanism. This would be consistent with the mechanism powering kilo-parsec and halo scale structures. However, the radio 
structure does extend beyond main [OIII] structure and alignments are not exact on the larger scale. In addition, as mentioned above, 
the continuum and emission lines show similar structures close to the nucleus. Coincident features such as these may be giving 
us supplementary clues about the nuclear outflow mechanisms. If we first naively assume that the continuum is dominated by stars and the 
emission lines are due mostly to gas, then it tells us the gas dynamics are currently consistent with the gravitational potential defined
by the stars. However, if the extended nuclear continuum is not stellar, but instead represents AGN-related continuum from the emission 
line regions, then we may be observing nebular continuum or scattered AGN light. Finally, the nuclear morphologies of the continuum and 
emission lines may be determined by uneven dust obscuration rather than AGN-related processes. This last case is considerably strong 
considering the evidence for clear dust lanes in both PKS 1549-79 and PKS 1345+12. 

Unfortunately, in the light of this, it is difficult to unambiguously determine whether radiative winds from the AGN, or outflows induced 
via the radio structures, dominate the driving mechanism. Further factors compound this problem. Even with the high spatial resolution of 
ACS, there is the fact that the resolution is unfavorable for clearly seeing detailed associations between radio and optical features. The 
problem is more acute if we consider the possible orientations of the small-scale bi-conical structures indicative of nuclear winds. There 
is some debate as to the relative alignment to the line of sight for the jet in PKS 1345+12, however, an orientation close to end on is 
favored for PKS 1549-79. In that case, the projection of the bi-cone would make resolving the feature especially difficult. 

\section{Conclusions}\label{cons}

We have resolved the emission line outflow regions in two compact radio sources (PKS 1549-79 and PKS 1345+12) using the high resolution 
channel of ACS. Through morphological comparisons with existing radio data we have investigated the possible driving mechanisms for the 
observed nuclear outflows. These sources are ideal for this study as they are well known radio galaxies in which nuclear outflows have 
been unequivocally detected. They also contain all of the possible driving mechanisms (jets, quasars and starbursts). 

In both cases we see good agreement with the radio and optical PAs and no evidence for the large scale bi-conical features 
indicative of starburst driven winds; the emission line morphology shows similar structure and scales to that of the radio. We are thus 
left with the conclusion that the outflows are most likely not driven by starburst winds. However, these ACS observations are unable to 
clearly resolve the scales over which a compact nuclear structures may exist, especially if orientations to the line-of-sight are 
considered. There is also an excellent agreement between the emission line and continuum morphology, which could be attributed to patchy 
dust obscuration. In addition, while the kinematics do advocate shocks, the ionization mechanism is difficult to couple with shock 
ionization. Radio jets and radiative AGN winds cannot be unambiguously distinguished as the dominant outflow mechanism.

\acknowledgments

We extend our thanks to Carlo Stanghellini, Istituto di Radioastronomia del C.N.R, Bologna, Italy, for supplying the 5 GHz radio data 
for PKS 1345+12, and to Tasso Tzioumis, CSIRO Radiophysics Laboratory, Pembroke and Vimiera Rds, Marsfield, NSW, Australia, for 
retrieving the radio data for PKS 1549-79 from Edward King's thesis. We also thank the anonymous referee for their useful comments. 

Support for Proposal number HST-GO-09401.10A was provided by NASA through a grant from the Space Telescope Science Institute, which is 
operated by the Association of Universities for Research in Astronomy, Incorporated, under NASA contract NAS5-26555.

\end{document}